# Fast core rotation in red-giant stars revealed by gravity-dominated mixed modes

Paul G. Beck[1], Josefina Montalban[2], Thomas Kallinger[1,3], Joris De Ridder[1], Conny Aerts[1,4], Rafael A. García[5], Saskia Hekker[6,7], Marc-Antoine Dupret[2], Benoit Mosser[8], Patrick Eggenberger[9], Dennis Stello[10], Yvonne Elsworth[7], Søren Frandsen[11], Fabien Carrier[1], Michel Hillen[1], Michael Gruberbauer[12], Jørgen Christensen-Dalsgaard[11], Andrea Miglio[7], Marica Valentini[2], Timothy R. Bedding[10], Hans Kjeldsen[11], Forrest R. Girouard[13], Jennifer R. Hall[13], Khadeejah A. Ibrahim[13]

1. Instituut voor Sterrenkunde, Katholieke Universiteit Leuven, 3001 Leuven, Belgium.
2. Institut d'Astrophysique et de Géophysique de l'Université de Liège, 4000 Liège, Belgium.
3. Institut für Astronomie der Universität Wien, Türkenschanzstraße 17, 1180 Wien, Austria.
4. Afdeling Sterrenkunde, IMAPP, Radboud University Nijmegen, 6500GL Nijmegen, The Netherlands.
5. Laboratoire AIM, CEA/DSM-CNRS-Université Paris Diderot; IRFU/SAp, Centre de Saclay, 91191 Gif-sur-Yvette Cedex, France.
6. Astronomical Institute 'Anton Pannekoek', University of Amsterdam, Science Park 904, 1098 XH Amsterdam, The Netherlands.
7. School of Physics and Astronomy, University of Birmingham, B15 2TT, Edgbaston, Birmingham UK.
8. LESIA, CNRS, Université Pierre et Marie Curie, Université Denis Diderot, Observatoire de Paris, 92195 Meudon Cedex, France.
9. Observatoire de Genève, Université de Genève, 51 Ch. des Maillettes, 1290 Sauverny, Switzerland.
10. Sydney Institute for Astronomy (SIfA), School of Physics, University of Sydney 2006, Australia.
11. Department of Physics and Astronomy, Aarhus University, DK-8000 Aarhus C, Denmark.
12. Department of Astronomy and Physics, Saint Marys University, Halifax, NS B3H 3C3, Canada.
13. Orbital Sciences Corporation/NASA Ames Research Center, Moffett Field, CA 94035, USA

**When the core hydrogen is exhausted during stellar evolution, the central region of a star contracts and the outer envelope expands and cools, giving rise to a red giant, in which convection occupies a large fraction of the star. Conservation of angular momentum requires that the cores of these stars rotate faster than their envelopes, and indirect evidence supports this[1,2]. Information about the angular momentum distribution is inaccessible to direct observations, but it can be extracted from the effect of rotation on oscillation modes that probe the stellar interior. Here, we report the detection of non-rigid rotation in the interiors of red-giant stars by exploiting the rotational frequency splitting of recently detected mixed modes[3,4]. We demonstrate an increasing rotation rate from the surface of the star to the stellar core. Comparing with theoretical stellar models, we conclude that the core must rotate at least ten times faster than the surface. This observational result confirms the theoretical prediction of a steep gradient in the rotation profile towards the deep stellar interior[1,5,6].**

The asteroseismic approach to studying stellar interiors exploits information from oscillation modes of different radial order n and angular degree ℓ, which propagate in cavities extending at different depths[7]. Stellar rotation lifts the degeneracy of non-radial modes, producing a multiplet of (2ℓ+1) frequency peaks in the power spectrum for each mode. The frequency separation between two mode components of a multiplet is related to the angular velocity and to the properties of the mode in its propagation region. More information on the exploitation of rotational splitting of modes may be found in the Supplementary Information. An important new tool comes from mixed modes which were recently identified in red giants[3,4]. Stochastically excited solar-like oscillations in evolved G- and K-giant stars[8] have been well studied in terms of theory[9-12], and the main results are consistent with recent observations from space-based photometry[13,14]. Unlike pressure modes, which are completely trapped in the outer acoustic cavity, mixed modes also probe the central regions and carry additional information from the core region, which is probed by gravity modes. Mixed dipole modes (l=1) appear in the Fourier power spectrum as dense clusters of modes around those of them best trapped in the acoustic cavity. These clusters, whose components contain a different amount of pressure and gravity mode influence, are referred to as 'dipole forests'.

We present the Fourier spectra of the brightness variations of KIC 8366239 (Fig. 1a), KIC 5356201 (Supplementary Fig. 3a) and KIC 12008916 (Supplementary Fig. 5a), derived from observations with the Kepler spacecraft. The three spectra show split modes, whose spherical degree we identify as $\ell=1$. We rule out finite mode lifetime effects through mode damping as a possible cause of these detected multiplets as they would not lead to a consistent multiplet appearance over several orders as shown in Fig. 1. The spacings in period between the multiplet components (Supplementary Fig. 7) are too small to be due to consecutive unsplit mixed modes[4] and do not follow the characteristic frequency pattern of unsplit mixed modes[3]. Finally, the projected surface velocity, $v \sin i$, obtained from ground-based spectroscopy (Tab.1), is consistent with the rotational velocity measured from the frequency splitting of the mixed mode that predominantly probe the outer layers. We are thus left with rotation as the only cause of the detected splittings.

The observed rotational splitting is not constant for consecutive dipole modes, even within a given dipole forest (Fig. 1b, and Supplementary Figs. 3b and 5b). The lowest splitting is generally present for the mode at the centre of the dipole forest, which is the mode with the largest amplitude in the outer layers, and it increases for modes with a larger gravity component towards the wings of the dipole mode forest. For KIC 8366239, we find that the average splitting of modes in the wings of the dipole forests is 1.5 times larger than the mean splitting of the centre modes of the dipole forests.

The observations (Fig. 1b) were confronted with theoretical predictions for a model representative of KIC 8366239 as defined in the Supplementary Information. The effect of rotation on the oscillation frequencies can be estimated in terms of a weight function, called a rotational kernel ($K_{nl}$). From the kernels it is shown that at least 60% of the frequency splitting for the $\ell=1$ mixed modes with a dominant gravity component is produced in the central region of the star (Fig. 2). This substantial core contribution to mixed modes enables us to investigate the rotational properties of the core region, which was hitherto not possible for the Sun, due to a lack of observed modes that probe the core region (within a radius $r < 0.2\ R_\odot$; Ref 15). The solar rotational profile is only known in great detail for those regions probed by pressure-dominated modes[16-18]. In contrast to these modes in the wings of the dipole forest, only 30% of the splitting of the centre mode originates from the central region of the star, while the outer third by mass of the star contributes 50% of the frequency splitting. By comparing the rotation velocity derived from the splitting of such pressure-dominated modes with the projected surface velocity from spectroscopy, we find that the asteroseismic value is systematically larger. This offset cannot be explained by inclination of the rotation axis towards the observer alone (Supplementary Table 1 and 2) but originates from the contribution of the fast rotating core (Fig. 2). Additionally, internal non-rigid rotation leads to a larger splitting for modes in the wings of the dipole forest than for centre modes. For a model rotating rigidly the reverse behaviour is expected (Fig. 1c). As the prediction for the rigidly rotating model is incompatible with the observed trends of the splittings (Fig. 1b) but can qualitatively be well reproduced under the assumption of non-rigid rotation (Fig. 1c), we conclude that the three stars investigated here (see Tab. 1, last column) rotate non-rigidly, with the central region rotating much faster than the surface.

The above interpretation is consistent with the correlation between the mode lifetime and the corresponding rotational splitting that has been observed in our data. Mixed modes in the wings of the dipole forest are predicted to have large amplitudes in the central regions of the star and, therefore, larger values of inertia and lifetime. These modes have narrower mode profiles in the frequency spectrum than the centre modes that are predominantly trapped in the outer cavity[11]. This behaviour of the mode profiles (Supplementary Fig. 1) was recently confirmed by observations[3,4]. We indeed observe that short-lived modes (with broader profiles) exhibit smaller rotational splitting in KIC 8366239 (Fig. 3) and the two other stars from Tab. 1 (Supplementary Figs. 9 and 10). With increasing lifetime (dominant gravity component) we see a substantial increase in the size of the rotational splitting of the modes. The frequencies of the narrowest modes are mainly affected by the rotation in the central regions of the star. Taking a representative model for KIC 8366239, and assuming that the convective envelope rotates rigidly and that the radiative interior rotates also rigidly but faster than the convective envelope, we find that the observed splitting ratio of 1.5 (Tab. 1) is obtained from a core rotating at least ten times faster than the surface.

So far our understanding of the evolution of the angular momentum as a function of the evolutionary stage is still poor. Previous to this study, it was only possible to deliver evidence of non-rigid rotation rate in two core hydrogen burning massive stars, for which the seismic analysis revealed a near-core rotation rate about three to five times faster than their envelope[19,20]. The low-luminosity red giants[21] discussed here are in the early phase of hydrogen-shell burning and have a core-to-envelope rotation rate more than twice as high as those of the massive stars (Ref 19,20). When these stars will enter the subsequent phase of core-helium burning, the core is believed to undergo a slight expansion, along with a shrinkage of the envelope, leading to an increased surface rotation on the horizontal branch compared to the one on the red giant branch[1]. The basic assumption of conservation of angular momentum predicts a less steep rotational gradient between surface and core than the one detected. Exploiting rotational splitting of mixed modes in a large sample of red giants in various evolutionary stages will provide an excellent tool to inspect how the internal angular momentum distribution evolves with time towards the end of the life of the star.

**Supplementary Information** is linked to the online version of the paper at www.nature.com/nature

**Acknowledgements**

We acknowledge the work of the team behind *Kepler*. Funding for the *Kepler* Mission is provided by NASA's Science Mission Directorate. P.B. and C.A. were supported by the European Community's 7'th Framework Programme (ERC grant PROSPERITY); J.D.R. and T.K. were supported by the FWO-Flanders. S.H. was supported by the Netherlands Organisation for Scientific Research (NWO). J.M. and M.V. were supported by the Belgian Science Policy Office. Based on observations with the HERMES spectrograph at the Mercator Telescope which is operated at La Palma/Spain by the Flemish Community.

**Authors Contribution**
P.G.B., T.K., J.D.R., C.A., R.A.G., S.H., B.M., Y.E., S.F., F.C., and M.G. measured the mode parameters, and derived and interpretated the rotational splitting and period spacings. J.M., M.-A.D., P.E., J.C.-D., A.M. calculated stellar models and provided theoretical interpretation of the rotational splitting. M.H., and V.A. observed and analyzed the spectra. J.D.R., S.H., S.F., Y.P.E., D.S., T.M.B., H.K., F.R.G., J.R.H., and K.A.I. contributed to the coordination of the project, including the acquisition and distribution of the data. C.A. defined and supervised the research. All authors discussed the results and commented on the manuscript.

**Author Information**
Reprints and permissions information is available at www.nature.com/reprints. The authors declare that they have no competing financial interests. Correspondence and requests for materials should be addressed to P.G.B. (e-mail: paul.beck@ster.kuleuven.be)

| KIC | $\nu_{max}$ [μHz] | R [$R_\odot$] | $\Delta P_{obs}$ [sec] | $T_{eff}$ [K] | $v \sin i$ [km/s] | Asteros. rot. vel. [km/s] | $\ell=1$ min rot. split. [μHz] | Averaged split. ratio for dipole modes |
|---|---|---|---|---|---|---|---|---|
| 8366239 | 182±1 | 5.30±0.08 | 56±11 | 4980±120 | < 1 | 3.9 | 0.135 ±0.008 | 1.5 |
| 5356201 | 209.7±0.7 | 4.47±0.03 | 50±10 | 4840±90 | 2.4 | 3.8 | 0.154 ±0.003 | 1.7 |
| 12008916 | 159.9±0.6 | 5.18±0.05 | 52±7 | 4830±100 | - | 5.7 | 0.200 ±0.001 | 1.8 |

**Table 1. Observational parameters of three stars, showing rotational splitting.** We present three cases of firm detections of rotational splitting in red giants from ~500 day long time series of photometric data obtained with the *Kepler* satellite in long-cadence mode (~30 min time sampling)[22]. The star's identifier in the *Kepler* Input Catalog (KIC) are given in the first column. The frequency of maximum oscillation power is given by $\nu_{max}$. The stellar radius R and the effective temperature $T_{eff}$ were computed according to Ref 14. The mean period spacing $\Delta P_{obs}$ and its standard deviation were derived from the central multiplet components of the dipole modes. These three stars are classified as low-luminosity red-giant stars[21] and the observationally derived mean period spacing of the mixed modes indicates that they are in an early phase of red-giant evolution[4], in which they are burning hydrogen in a shell around the helium core. The projected surface velocity $v \sin i$ was determined from ground-based spectra. The asteroseismic rotation velocity was computed from the minimum dipole splitting and assuming rigid rotation. Although this is the most pressure-dominated mode found in the spectrum, it contains a contribution from the core and the inferred rotation velocity from this mode is therefore slightly higher than the surface rotation velocity. The observed ratio of the average rotational splitting found in the wings of the dipole forest to the mean splitting of the forest centre is given in the last column. Extensive lists of further asteroseismic parameters are given in Supplementary Table 1 and 2.

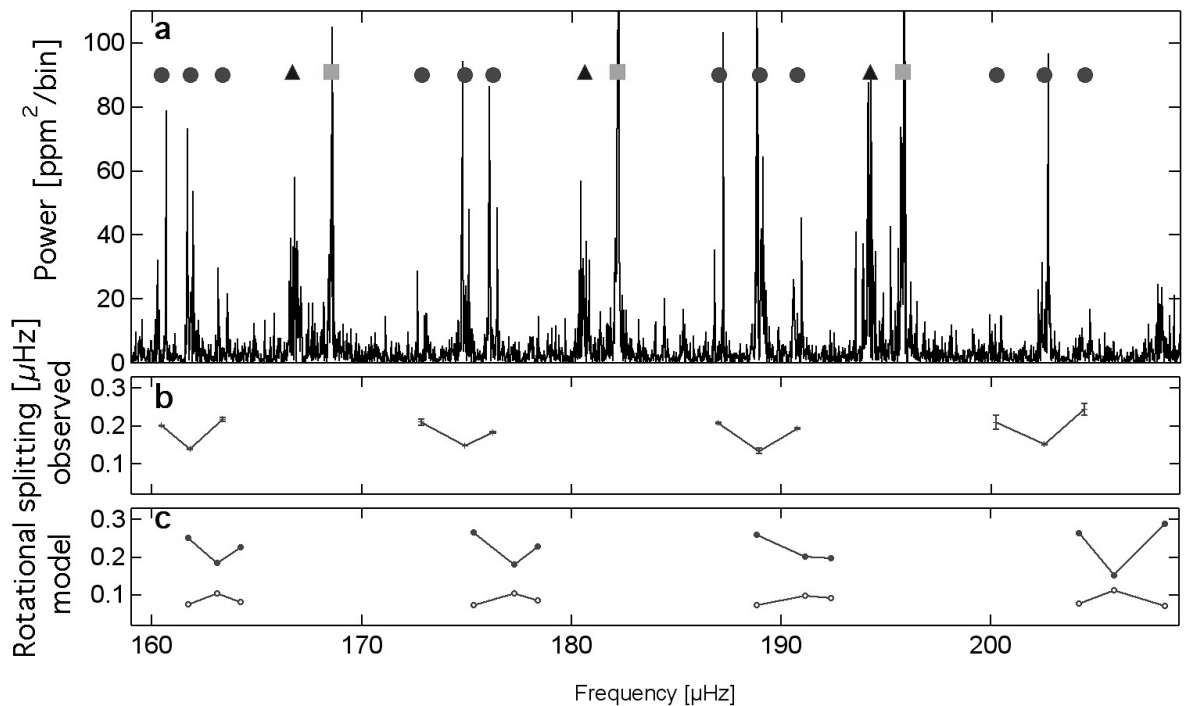

**Figure 1 | Oscillation spectrum of KIC 8366239. a**, The radial and $\ell=2$ modes are marked with squares and triangles, respectively. Each $\ell=1$ rotational multiplet is marked with one dot. A zoom on the region of 185 to 195 µHz and the analysis of the comb-like structure of the oscillation spectrum in a so-called échelle diagram are shown in the online material (Supplementary Figs. 1 and 2). The spectral window of the Fourier analysis can be found in the online material (Supplementary Fig. 11). **b,** The observed rotational splitting for individual $\ell=1$ modes. The error bars indicate the standard deviation of the measured rotational splitting of dipole modes. Similar analyses of the stars KIC 5356201 (Supplementary Figs. 3 and 4) and KIC 12008916 (Supplementary Figs. 5 and 6) are discussed in the online material. **c,** Theoretically predicted rotational splitting assuming two different rotation laws. The values are calculated for a representative model of KIC 8366239 as defined in the Supplementary Information. The splitting for nonrigid rotation (marked with solid dots) for the case of a core rotation 10 times faster than the surface rotation of 2.5 km/s resembles the observations qualitatively well. The theoretical splittings for the scenario of rigid rotation and an equatorial surface rotation velocity of 3 km/s (marked with open symbols) show a trend opposite to the observed one, with the largest splitting in the centre of the dipole forest and lower splitting in gravity-dominated modes. In the case of rigid rotation, the variable splitting is purely governed by the variation of the Ledoux[23] constant across the dipole forest (Supplementary Fig. 8). As the representative model (in panel c) has not been corrected for surface effects, there is a slight offset to the observations (in panel b). (Coloured symbols are shown in the online version of the paper.)

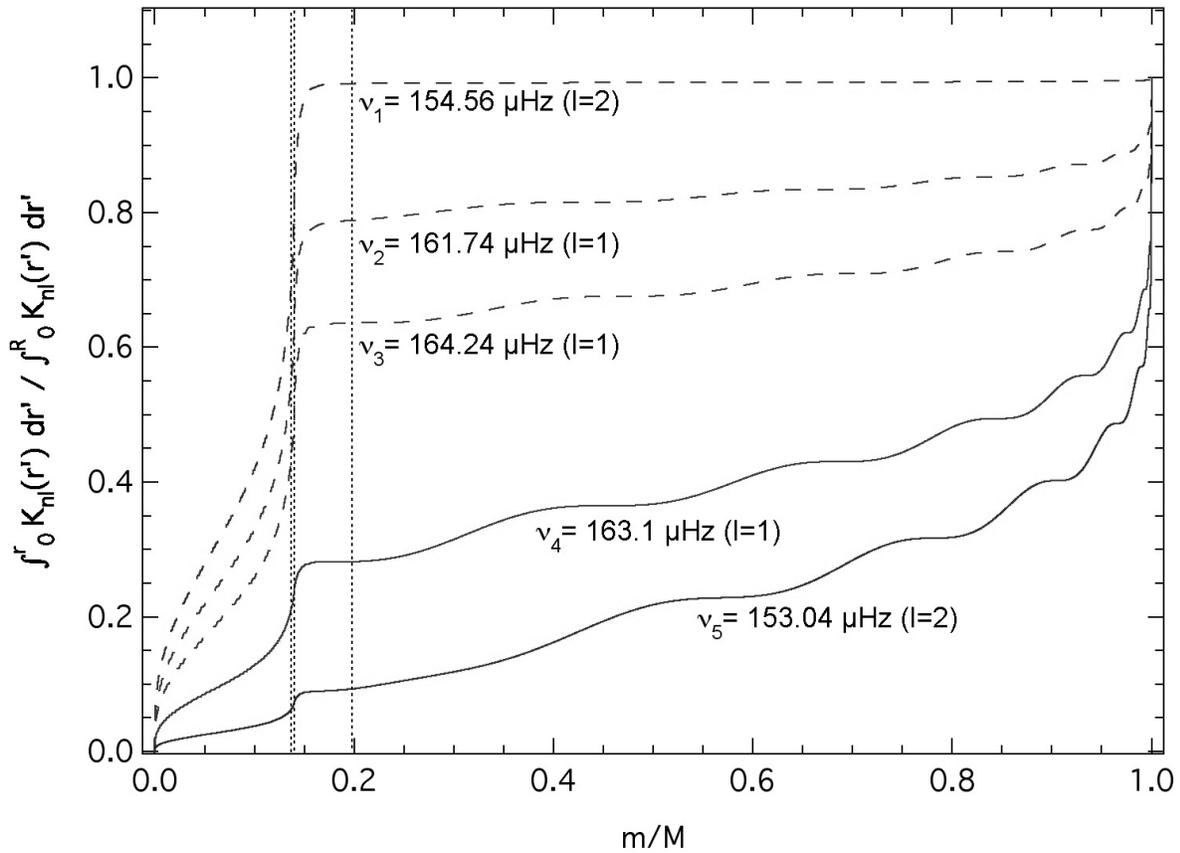

**Figure 2 | Contributions to the total rotational splitting.** Partial integrals of normalized rotation kernels, illustrating the contribution from different regions to the rotational splitting for pressure-dominated modes ($\nu_1$, $\nu_2$ shown with solid lines) and gravity-dominated modes ($\nu_2$, $\nu_3$, $\nu_5$ shown with dashed lines) of degree $\ell=1$ and $\ell=2$, as a function of the stellar mass-fraction. The kernels have been calculated for modes from a representative model of KIC 8366239, as defined in the Supplementary Information, with oscillation frequencies at $\nu_1$=154.56 µHz, $\nu_2$=161.74 µHz, $\nu_3$=164.24 µHz, $\nu_4$=163.10 µHz, , and $\nu_5$=153.04 µHz. Vertical dotted lines correspond, from left to right, to the boundary of the helium core, the H-burning shell and the bottom of the convective envelope. (Coloured symbols are shown in the online version of the paper.)

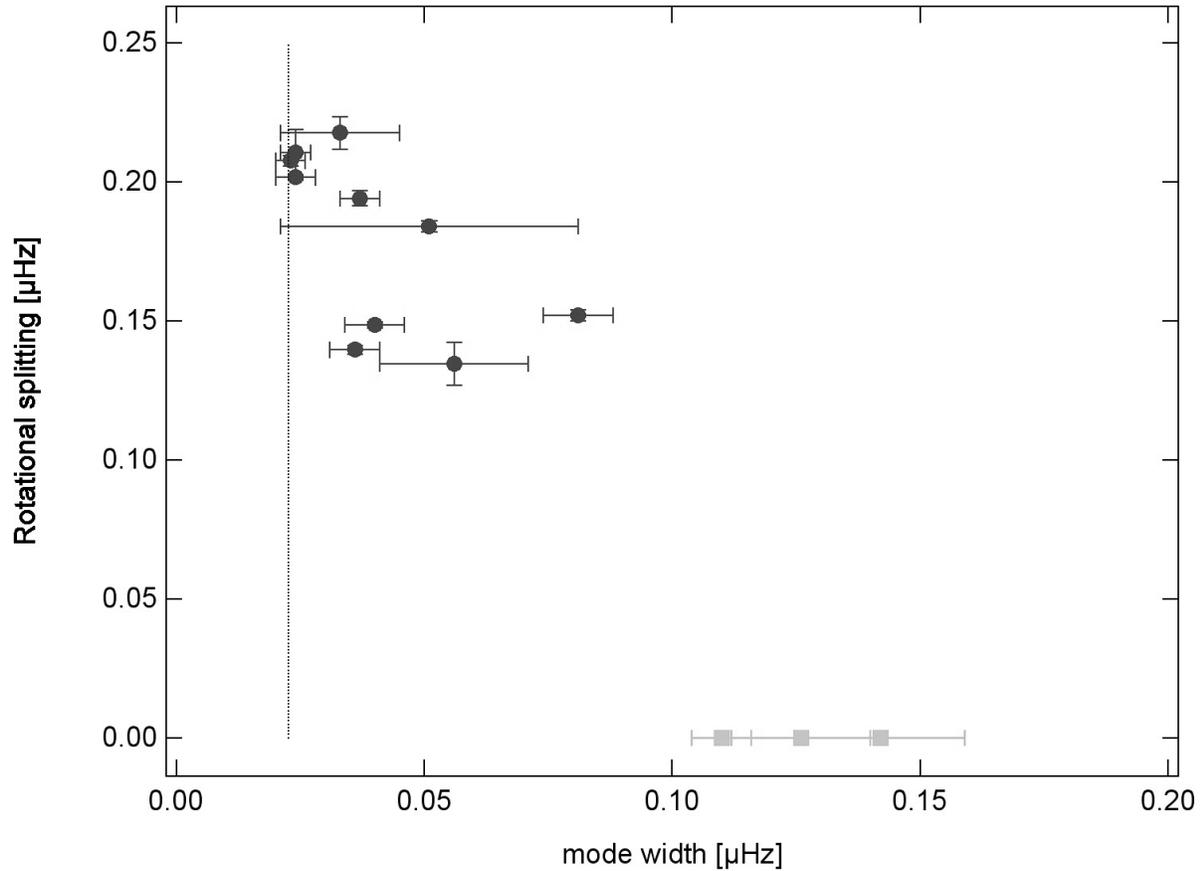

**Figure 3 | Rotational splitting versus mode linewidth for KIC 8366239.** The lifetime is inversely proportional to the mode linewidth. Similar to Fig. 1, ℓ=1 modes are marked with dots. For comparison also the linewidth of the pure acoustic radial modes (ℓ=0) are shown with squares. The dotted vertical line represents the formal frequency resolution. The error bars indicate the standard deviation of the measured rotational splitting and modewidth of dipole modes. Similar diagram for the two other stars from Tab. 1 are shown in Suppl. Fig 9 and 10, respectively. (Coloured symbols are shown in the online version of the paper.)